\documentclass[a4paper, 11pt]{article}\usepackage[]{graphicx}\usepackage[dvipsnames,table]{xcolor}
\makeatletter
\def\maxwidth{ %
  \ifdim\Gin@nat@width>\linewidth
    \linewidth
  \else
    \Gin@nat@width
  \fi
}
\makeatother

\definecolor{fgcolor}{rgb}{0.345, 0.345, 0.345}

\usepackage{framed}
\makeatletter
 {\par\unskip\endMakeFramed%
 \at@end@of@kframe}
\makeatother

\definecolor{shadecolor}{rgb}{.97, .97, .97}
\definecolor{messagecolor}{rgb}{0, 0, 0}
\definecolor{warningcolor}{rgb}{1, 0, 1}
\definecolor{errorcolor}{rgb}{1, 0, 0}
\newenvironment{knitrout}{}{} 

\usepackage{alltt}
\usepackage{amsmath, amssymb}
\usepackage{doi} 
\usepackage[round]{natbib} 
\usepackage{multirow}
\usepackage{booktabs} 
\usepackage[title]{appendix} 
\usepackage{nameref} 
\usepackage[dvipsnames,table]{xcolor}
\usepackage[doublespacing]{setspace} 
\usepackage[labelfont=bf,font=small]{caption} 
\usepackage{helvet}
\usepackage{mathpazo}
\usepackage{sectsty} 
\allsectionsfont{\sffamily} 
\usepackage[labelfont={bf,sf},font=small]{caption} 
\usepackage{orcidlink} 

\usepackage{geometry}
\newif\ifanonymize 
\newcommand{\anonymize}[1]{%
  \ifanonymize
    \phantom{#1}%
  \else
    #1%
  \fi
}
\anonymizefalse 

\newcommand\mail{samuel.pawel@uzh.ch}
\title{\vspace{-2em}
  \textbf{\textsf{A Bayes Factor Framework for Unified Parameter Estimation and Hypothesis Testing}}
}
\author{
 \anonymize{\textbf{Samuel Pawel} \orcidlink{0000-0003-2779-320X}} \\
  \anonymize{Epidemiology, Biostatistics and Prevention Institute (EBPI)} \\
  \anonymize{Center for Reproducible Science (CRS)} \\
  \anonymize{University of Zurich} \\
  \anonymize{E-mail: \href{mailto:\mail}{\mail}} \\[2ex]
  \anonymize{{\color{blue} Preprint version July 8, 2025}}
}
\date{}

\usepackage{hyperref}
\hypersetup{
  bookmarksopen=true,
  breaklinks=true,
  colorlinks=true,
  linkcolor=blue,
  anchorcolor=black,
  citecolor=blue,
  urlcolor=black,
}

\usepackage{fancyhdr}
\DeclareMathOperator{\BF}{BF} 
\newcommand{\that}{\hat{\theta}} 
\DeclareMathOperator*{\argmax}{arg\,max} 
\DeclareMathOperator{\Nor}{N} 
\DeclareMathOperator{\Bin}{Bin} 
\DeclareMathOperator{\Beta}{Beta} 
\newcommand{\B}{\operatorname{{B}}} 
\newcommand{\thatME}{\that_{\mathrm{ME}}} 
\newcommand{\kME}{k_\mathrm{ME}} 

\IfFileExists{upquote.sty}{\usepackage{upquote}}{}
\begin{document}

\maketitle

\begin{abstract}
  The Bayes factor, the data-based updating factor of the prior to posterior
  odds of two hypotheses, is a natural measure of statistical evidence for one
  hypothesis over the other. We show how Bayes factors can also be used for
  parameter estimation. The key idea is to consider the Bayes factor as a
  function of the parameter value under the null hypothesis. This `support
  curve' is inverted to obtain point estimates (`maximum evidence estimates')
  and interval estimates (`support intervals'), similar to how \textit{P}-value
  functions are inverted to obtain point estimates and confidence intervals.
  This provides data analysts with a unified inference framework as Bayes
  factors (for any tested parameter value), support intervals (at any level),
  and point estimates can be easily read off from a plot of the support curve.
  This approach shares similarities but is also distinct from conventional
  Bayesian and frequentist approaches: It uses the Bayesian evidence calculus,
  but without synthesizing data and prior, and it defines statistical evidence
  in terms of (integrated) likelihood ratios, but also includes a natural way
  for dealing with nuisance parameters. Applications to meta-analysis,
  replication studies, and logistic regression illustrate how our framework is
  of practical value for making quantitative inferences. \\[1ex]
\end{abstract}

\section{Introduction}

A universal problem in data analysis is making inferences about unknown
parameters of a statistical model based on observed data. In practice, data
analysts are often interested in two tasks: (i) estimating the parameters (i.e.,
finding the most plausible value or a region of plausible values based on the
observed data), and (ii) testing hypotheses related to them (i.e., using the
observed data to quantify the evidence that the parameter takes a certain
value). While these tasks may seem distinct, there are several statistical
concepts that provide a link between the two.

\begin{figure}[!htb]
\begin{knitrout}
\definecolor{shadecolor}{rgb}{0.969, 0.969, 0.969}\color{fgcolor}

{\centering \includegraphics[width=\maxwidth]{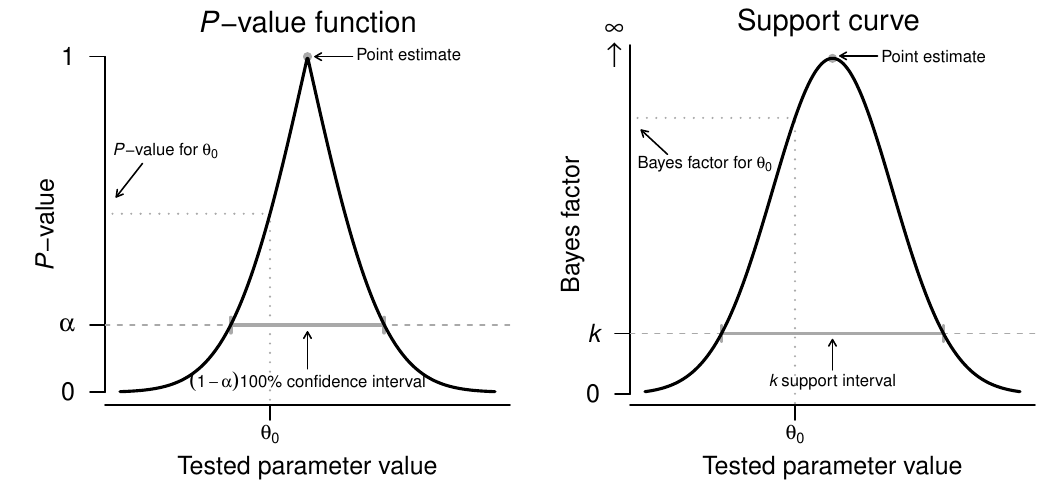} 

}

\end{knitrout}
\caption{Examples of \textit{P}-value functions and support curves.
  \textit{P}-value are two-sided. Bayes factors are oriented in favour of the
  tested parameter value over a specified alternative hypothesis (i.e., a higher
  Bayes factor indicates higher support for the parameter value over the
  alternative).}
\label{fig:pvalfun}
\end{figure}

In frequentist statistics, there is a duality between parameter estimation and
hypothesis testing as \textit{P}-values, confidence intervals, and point
estimates correspond in the sense that the \textit{P}-value for a tested
parameter value is less than $\alpha$ if the $(1-\alpha)100\%$ confidence
interval excludes that parameter value, and that the (two-sided)
\textit{P}-value is largest when the tested parameter value is the point
estimate. The \emph{\textit{P}-value function} -- the \textit{P}-value viewed as
a function of the tested parameter \citep[for an overview see
e.g.,][]{Bender2005, Fraser2019} -- provides a link between these concepts. One
may alternatively look at closely related quantities: One minus the two-sided
\textit{P}-value function known as \emph{confidence curve} \citep{Cox1958b,
  Birnbaum1961}, one minus the one-sided \textit{P}-value function known as
\emph{confidence distribution}, or its derivative known as \emph{confidence
  density} \citep{XieSingh2013, SchwederHjort2016}. A visualization of the
\textit{P}-value function, such as shown in the left plot in
Figure~\ref{fig:pvalfun}, provides the observer with a wealth of information, as
\textit{P}-values (for any tested parameter), confidence intervals (at any level
of interest), and point estimates can be easily read off. As such,
\textit{P}-value functions and their relatives have been deemed important
measures to address common misinterpretations and misuses of \textit{P}-values
and confidence intervals \citep[among others]{Greenland2016, Infanger2019,
  Rafi2020, Marschner2024}.

In Bayesian statistics, the posterior distribution of the unknown parameter
plays a similar role to the \textit{P}-value function, since point estimates
(e.g., posterior modes, medians, or means), credible intervals, and posterior
probabilities of hypotheses can all be derived from it. The posterior provides a
synthesis of the data and the prior distribution, which can be seen as an
advantage but also as a challenge in the absence of prior knowledge. In
particular, for testing of hypotheses, it can be difficult to specify prior
probabilities such as `$\Pr(\text{the treatment effect is absent})$' and
`$\Pr(\text{the treatment effect is present})$'.

\begin{sloppypar}
One approach to address this issue is to report the \emph{Bayes factor}
\citep{Jeffreys1939, Good1958, Kass1995}, i.e., the updating factor of the prior
to posterior odds of two hypotheses. As such, Bayes factors allow data analysts
to evaluate the relative evidence for two hypotheses without depending on the
prior probabilities of the hypotheses; for example, a Bayes factor can quantify
the evidence for the presence or absence of a treatment effect without having to
assign prior probabilities to these hypotheses (although one still has to
specify a prior for the parameter under the alternative, which is challenging in
itself). However, the use of Bayes factors comes at the cost of lacking an
overarching concept, such as a \textit{P}-value function or posterior
distribution, that can provide data analyst with a coherent set of point and
interval estimates. In practice, data analysts who wish to perform hypothesis
testing with Bayes factors but also parameter estimation are therefore faced
with a dilemma; they can either supply their Bayes factors with a posterior
distribution conditional on one hypothesis being true (e.g., the posterior of a
treatment effect, assuming the effect is indeed present), which can lead to
contradictory conclusions with the Bayes factor \citep[for examples,
  see][]{Stone1997, Wagenmakers2020, Kelter2022}, or they can assign prior
probabilities to the tested hypotheses and report a posterior averaged over both
hypotheses \citep[known as Bayesian model averaging, see
  e.g.,][]{Hoeting1999,Rouder2018b,Campbell2022}, but this requires
specification of prior probabilities which is highly controversial and usually
the reason why the Bayes factor was reported in the first place rather than the
posterior probabilities of the hypotheses.
\end{sloppypar}

Our goal is therefore to resolve this dilemma and provide a unified framework
for estimation and hypothesis testing based on Bayes factors. The idea is the
same as for the \textit{P}-value function; we consider the Bayes factor as a
function of the tested parameter. We then use this \emph{support curve} to
derive point estimates, interval estimates, and Bayes factors (as shown in the
right plot in Figure~\ref{fig:pvalfun}), establishing a duality between
hypothesis testing and parameter estimation. This provides data analysts with a
unified framework for statistical inference centred around the Bayes factor.

Interval estimates based on the Bayes factor have been proposed under the name
of the Bayesian support interval \citep{Wagenmakers2020, Pawel2023}. Similarly,
looking at the Bayes factor as a function of parameter values has been proposed
in the physics community under the names of `Bayes factor surface'
\citep{Fowlie2024} and `\textit{K} ratio' \citep{Afzal2023}, and by
\citet{Rouder2018} in the context of teaching Bayesian statistics. A method
called `Bayes factor function' has been proposed by \citet{Johnson2023}. In this
approach, Bayes factors are viewed as a function of a hyperparameter of the
prior \emph{under the alternative hypothesis} but for a fixed null hypothesis.
The Bayes factor function can in some cases be made equivalent to the support
curve if the roles of the null and alternative hypothesis are reversed and point
mass priors are assigned to the parameter under the alternative, yet this seems
rather unnatural from the perspective of both methods. Finally, a similar
attempt of reconciling Bayesian interval estimation and hypothesis testing with
a focus on the `Bayesian evidence value' has been investigated by
\citet{Kelter2022}.

While many elements of support curve inference have been considered in earlier
articles, this paper provides a synthesis of these ideas from the perspective of
the support curve as a counterpart to the \textit{P}-value function. A novel
contribution is the concept of point estimation based on Bayes factors. We call
the resulting estimate the \emph{maximum evidence estimate} (MEE). The MEE is
the parameter value that maximizes the support curve, and as such, is the
parameter value that receives the most evidential support from the data over a
specified alternative hypothesis. Finally, this paper shows how support curves
can be applied in several areas where they have never been used before
(meta-analysis, replication studies, and logistic regression).

It is important to note that our intent is not to fuel the `statistics wars' and
entertain the position that Bayesian inference or Bayes factors are the best or
only valid form of statistical inference. The goal is simply to explore the idea
of a Bayes factor analogue of the \textit{P}-value function, find out how such
an approach can be useful, and how it connects to other approaches. Bayes
factors, just as \textit{P}-values and posterior probabilitites, can be delicate
tools and users should be wary of their limitations, misuses, and
misinterpretations \citep{Tendeiro2019, Wong2022, Tendeiro2024}.

This paper is structured as follows. In the following
(Section~\ref{sec:BFtheory}), we introduce the theoretical foundations of Bayes
factors, support sets, and maximum evidence estimates. We then explore their
connection to other statistical frameworks (Section~\ref{sec:connections}). Real
data examples from meta-analysis, replication studies, and logistic regression
then illustrate properties of the Bayes factor framework
(Section~\ref{sec:applications}). We conclude with a discussion of limitations,
advantages, and directions for future research (Section~\ref{sec:discussion}).

\section{Support curve inference}
\label{sec:BFtheory}

Suppose we observe data $y$ with an assumed distribution with probability
density/mass function $p(y \mid \theta, \psi)$ that depends on parameters
$\theta \in \Theta$ and $\psi \in \Psi$, with $\theta$ being the focus
parameters (the parameters of substantial interest) and $\psi$ being possible
nuisance parameters. Consider two hypotheses, the null hypothesis $H_0 \colon
\theta = \theta_0$ postulating that $\theta$ takes a certain value $\theta_{0}$
and the alternative hypothesis $H_1 \colon \theta \neq \theta_0$ postulating
that $\theta$ takes a different value. A natural measure of relative evidence
for the two hypotheses is the Bayes factor \citep{Jeffreys1939, Good1958,
  Kass1995}, the data-based updating factor of the prior odds of the hypotheses
to their posterior odds
\begin{subequations}
\begin{align}
    \BF_{01}(y;\theta_0)
    &= \frac{p(H_0 \mid y)}{p(H_1 \mid y)} \bigg/ \, \frac{p(H_0)}{p(H_1)}
    \label{eq:bf01update} \\
    &= \frac{p(y \mid H_0)}{p(y \mid H_1)} \label{eq:bf01predictive} \\
    &= \frac{\int_{\Psi} p(y \mid \theta_0, \psi) p(\psi \mid H_0) \, \mathrm{d}\psi}{\int_{\Theta} \int_{\Psi} p(y \mid \theta, \psi) \,p(\theta, \psi \mid H_1) \,\mathrm{d}\psi \, \mathrm{d}\theta}
    \label{eq:bf01compute}
\end{align}
\label{eq:bf01}
\end{subequations}
with $p(\theta, \psi \mid H_1)$ denoting the prior assigned to the parameters
under $H_1$ and $p(\psi \mid H_0)$ the prior assigned to the nuisance parameters
under $H_0$. Note that Bayes factors do not require the null and alternative
models to be nested but are applicable to general model comparisons. However,
due to their more intuitive interpretation and computational advantages, we will
focus on nested models and discuss extensions to non-nested models in
Section~\ref{sec:discussion}.

As~\eqref{eq:bf01update} shows, the Bayes factor represents the data-based core
of the Bayesian belief calculus since it transforms the prior odds of $H_0$
versus $H_1$ to the corresponding posterior odds \citep{Goodman1999}. The
alternative expression of the Bayes factor in equation~\eqref{eq:bf01predictive}
shows that this update is dictated by the relative predictive accuracy of the
two hypotheses. That is, the posterior odds of the null hypothesis $H_{0}$
increase if it outperforms the competing alternative hypothesis $H_{1}$ in
predicting the data $y$, and vice versa \citep{Good1952, Gneiting2007}. Since
the Bayes factor remains the same for any pair of prior hypothesis probabilities
$p(H_0)$ and $p(H_1)$, it remains useful to different analysts with different
prior hypothesis probabilities, or even if one rejects the idea of assigning
probabilities to $H_0$ and $H_1$. However, this does not mean that the Bayes
factor does not depend on any prior distributions; in fact, it can be quite
sensitive to the priors assigned to the parameters $\psi$ (and $\theta$) under
$H_0$ and $H_1$, respectively \citep{Kass1995}. We will return to the choice of
parameter priors in Section~\ref{sec:priorchoice}. Finally, the last
equation~\eqref{eq:bf01compute} shows how the Bayes factor can be calculated,
i.e., by dividing the likelihood of $y$ under the null value $\theta_0$
(possibly integrated over the prior of $\psi$ under $H_0$) by the likelihood of
$y$ integrated over the prior of $\theta$ (and possibly $\psi$) under~$H_1$. The
priors for $\theta$ and $\psi$ may also be point mass priors, in which case the
Bayes factor reduces to a likelihood ratio.

The idea now is to consider the Bayes factor~\eqref{eq:bf01} as a function of
$\theta_0$, that is, to vary the tested parameter value (the null hypothesis
$H_0 \colon \theta = \theta_0$) in order to assess the support for different
parameter values over the alternative $H_{1}$, see the right plot in
Figure~\ref{fig:pvalfun} for an example. Like the \textit{P}-value function,
this \emph{support curve} (SC) helps to address cognitive challenges with
inferential statistics \citep{Greenland2017}. For example, it shifts the focus
of inference from testing a single privileged null hypothesis (e.g., the
hypothesis that there is no treatment effect) to an entire continuum of
hypotheses. By looking at the SC, data analysts can then identify hypotheses
that receive equal or even less support from the data than the privileged one;
for example, a parameter value indicating a very large treatment effect may
receive equal support as the value of no treatment effect \citep[sometimes
  called `counternull', see][]{Rosenthal1994, Bind2024}.

For one- or two-dimensional focus parameters $\theta$, the SC can be plotted as
a curve or surface, respectively, so that the relative support for parameter
values can be visually assessed. For higher dimensional focus parameters (for
example, more than two regression coefficients), this becomes more difficult. In
this case, some focus parameters can be re-considered as nuisance parameters to
reduce the SC visualization again to a curve or surface. This is similar to
ordinary Bayesian inference where a marginal posterior is obtained from
marginalizing a joint posterior.

\subsection{Support sets}
Apart from visual inspections, we may also want to summarize the SC numerically.
The SC can be used to obtain \emph{support sets} \citep{Wagenmakers2020} which
are set-valued estimates for $\theta$ based on inverting the Bayes
factor~\eqref{eq:bf01} similar to how \textit{P}-value functions are inverted to
obtain confidence sets. Specifically, a support set at support level~$k > 0$ is
defined by
\begin{align*}
    S_k = \left\{\theta_0 : \BF_{01}(y;\theta_0) \geq k\right\}
\end{align*}
that is, the parameter values for which the Bayes factor indicates evidence of
at least level~$k$ over the specified alternative. A $k$ support set thus
contains all the parameter values that, when considered as individual null
hypotheses compared to an alternative, receive a Bayes factor of at least $k$.
Each included parameter value is therefore associated with an updating factor of
at least $k$ or, equally, relative predictive performance of at least $k$.
Support sets are thus `calibrated' with respect to prior-posterior updating, as
opposed to the calibration of confidence sets with respect to long-run coverage,
or that of posterior credible intervals with respect to posterior belief.

In practice, a $k$ support set (typically an interval) is obtained from
`cutting' the SC at $k$ and taking the parameter values above as part of the
support set (see the right plot in Figure~\ref{fig:pvalfun} for illustration).
It may happen that for certain choices of $k$ the support set is empty because
the data do not constitute relative evidence at that level. The choice of the
support level is arbitrary, just as the choice of the confidence level from a
confidence set is. One may, for example, report the support level $k = 1$ as it
represents the tipping point at which the parameter values begin to be supported
over the alternative. Conventions for Bayes factor evidence levels can also be
used. For example, based on the convention from \citet{Jeffreys1961}, a support
set at level $k = 10$ includes the parameter values that receive `strong'
relative support from the data, while a $k = 1/10$ support set includes the
parameter values that are at least not strongly contradicted.

\subsection{The maximum evidence estimate}
A natural point estimate for the unknown parameter $\theta$ based on the SC is
given by
\begin{align*}
    \thatME = \argmax_{\theta_0 \in \Theta}  \BF_{01}(y;\theta_0),
\end{align*}
and we call it the \emph{maximum evidence estimate} (MEE), since it is the
parameter value for which the Bayes factor indicates the most evidence over the
alternative. The associated \emph{evidence level}
\begin{align*}
\kME = \BF_{01}(y;\thatME),
\end{align*}
that is, the SC evaluated at the MEE, quantifies the evidential value of the
estimate $\thatME$ over the alternative. Evidence levels close to $\kME = 1$
indicate that the MEE receives little support over the alternative hypothesis
$H_1$, whereas large evidence levels $\kME$ indicate that the MEE receives
substantial support over the alternative hypothesis $H_1$. A useful summary of a
SC is hence given by the MEE, its evidence level, and a support set, similar to
how a \textit{P}-value function may be summarized with a point estimate and
confidence set.

To understand the behaviour of the MEE with increasing sample size, we may look
at an approximation of the Bayes factor. Suppose that the data
$y_{1:n} = \{y_{1}, y_{2}, \dots, y_{n}\}$ are independent and identically
distributed and denote by $\hat{\psi}_{0}$ the maximizer of the log likelihood
of the data under the null and by $(\hat{\theta}_{1}$, $\hat{\psi}_{1})$ the
maximizer under the alternative hypothesis. Denote by $n \widehat{V}_{0}$ and
$n \widehat{V}_{1}$ the modal dispersion matrices (minus the inverse of the
matrix of second-order partial derivatives of the log likelihood evaluated at
the corresponding maximizer). Applying a Laplace approximation to the logarithm
of the SC \cite[equation 7.27]{OHagan2004} gives then
\begin{align}
  \log \BF_{01}(y_{1:n};\theta_{0}) \approx
  \log \frac{p(y_{1:n} \mid \theta_{0}, \hat{\psi}_{0})}{
  p(y_{1:n} \mid \hat{\theta}_{1}, \hat{\psi}_{1})}
  + \frac{\dim(\theta)}{2} \log \frac{n}{2\pi} +
  \log \frac{p(\hat{\psi}_{0} \mid H_{0})}{
  p(\hat{\theta}_{1}, \hat{\psi}_{1} \mid H_{1})}
  + \frac{1}{2} \log \frac{|\widehat{V}_{0}|}{|\widehat{V}_{1}|}.
  \label{eq:bfapprox}
\end{align}
To obtain the MEE, the log Bayes factor~\eqref{eq:bfapprox} needs to be
maximized with respect to $\theta_{0}$. It is clear that as $\theta_{0}$ becomes
more different from $\that_{1}$, the log normalized profile likelihood (first
term) will decrease toward negative infinity, indicating evidence against the
parameter value $\theta_{0}$. On the other hand, when $\theta_{0}$ is not too
far from $\that_{1}$ the term will be about zero, so that an increasing sample
size $n$ (second term) increases the log SC toward positive infinity,
indicating evidence for $\theta_{0}$. The relative accuracy of the priors (third
term) and the relative dispersion (fourth term) lead to further adjustments of
the SC. For instance, when a parameter estimate is likely under the
corresponding prior, this increases the evidence for corresponding hypothesis
while a misspecified prior that is in conflict with the parameter estimates
lowers the evidence for the corresponding hypothesis. In sum, finding the MEE
corresponds approximately to maximizing the profile likelihood that is adjusted
based on the accuracy of the prior of the parameters and the modal dispersion.

\subsection{Example: Normal mean}
\label{sec:normalmean}

Suppose we observe a mean $\bar{y}$ from $n$ observations, assumed to be sampled
(at least approximately) from a normal distribution $\bar{Y} \mid \theta \sim
\Nor(\theta, \kappa^{2}/n)$. Assume that the variance $\kappa^{2}$ is known and
we want to conduct inferences regarding $\theta$. For example, suppose we
observed a mean test score of $\bar{y} = 110$ in a sample of $n =
30$ students. Also assume that in the general population the average
score is 100 and the standard deviation is $\kappa = 15$. To obtain a
Bayes factor for contrasting $H_{0} \colon \theta = \theta_{0}$ against $H_{1}
\colon \theta \neq \theta_{0}$ we need to formulate a prior for $\theta$ under
the alternative $H_{1}$. We will now discuss three choices with different
characteristics shown in Table~\ref{tab:normalinference}.

\begin{figure}[!htb]
\begin{knitrout}
\definecolor{shadecolor}{rgb}{0.969, 0.969, 0.969}\color{fgcolor}
\includegraphics[width=\maxwidth]{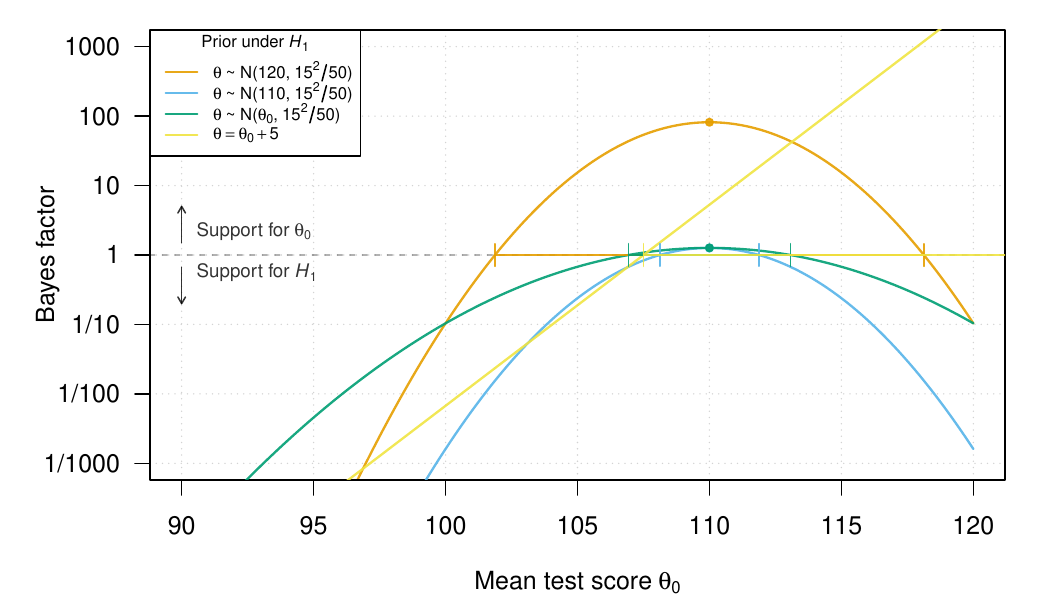} 
\end{knitrout}
\caption{Support curve, MEE, and $k = 1$ support interval for a mean
  test score $\theta$ based on the observed mean $\bar{y} =110$
  from $n = 30$ observations for different prior distributions for
  $\theta$ under the alternative $H_{1}$. A normal likelihood $\bar{Y} \mid
  \theta \sim \Nor(\theta, 15^2/n)$ is assumed for the data.}
\label{fig:normal}
\end{figure}

\begingroup
\renewcommand{\arraystretch}{2.25}
\begin{table}[!htb]
  \centering
  \caption{Support curve $\BF_{01}$, maximum evidence estimate $\thatME$,
    evidence value $\kME$, and $k$ support interval (SI) for a normal mean based
    on an observed mean $\bar{y}$ from $\bar{Y} \mid \theta \sim \Nor(\theta,
    \kappa^{2}/n)$ with known variance $\kappa^{2}$ and for three prior
    distributions for $\theta$ under the alternative $H_{1}$: A normal prior
    (left), a normal prior centered around the parameter value of the null
    hypothesis $\theta_{0}$ (middle), a point prior shifted away from the
    parameter value of the null hypothesis $\theta_{0}$ by $d > 0$ (right).}
  \label{tab:normalinference}
  \resizebox{\columnwidth}{!}{%
  \begin{tabular}{l c c c}
    \toprule
    & \multicolumn{3}{c}{Prior for $\theta$ under $H_{1}$} \\
    \cmidrule(lr){2-4}
    & $\theta \sim \Nor(m, v)$ & $\theta \sim \Nor(\theta_{0}, v)$ & $\theta = \theta_{0} + d$\\
    \midrule
    $\BF_{01}$ & $\exp\left[-\frac{1}{2}\left\{\frac{(\bar{y} - \theta_{0})^{2}}{\kappa^{2}/n} - \frac{(\bar{y} - m)^{2}}{v +\kappa^{2}/n}  \right\}\right] \sqrt{1 + \frac{v\, n}{\kappa^2}}$ & $\exp\left[-\frac{1}{2}\left\{\frac{n(\bar{y} - \theta_{0})^{2}}{\kappa^{2}\{1 + \kappa^{2}/(v\,n)\}}\right\}\right] \sqrt{1 + \frac{v\,n}{\kappa^2}}$ & $\exp\left\{\frac{2 d (\theta_{0} - \bar{y}) + d^{2}}{2 \kappa^{2}/n}\right\}$\\
    $\thatME$ & $\bar{y}$ & $\bar{y}$ & non-existent \\
    $\kME$ & $\exp\left\{\frac{(\bar{y} - m)^{2}}{2(v + \kappa^{2}/n)}  \right\} \sqrt{1 + \frac{v\, n}{\kappa^2}}$ & $\sqrt{1 + \frac{v\, n}{\kappa^2}}$ & non-existent \\
    $k$ SI & $\bar{y} \pm \frac{\kappa}{\sqrt{n}} \sqrt{\log(1 + \frac{v\, n}{\kappa^2}) + \frac{(\bar{y} - m)^{2}}{v + \kappa^{2}/n} - \log k^{2}}$ & $\bar{y} \pm \frac{\kappa}{\sqrt{n}} \sqrt{\left\{\log(1 + \frac{v \, n}{\kappa^2}) - \log k^{2}\right\}(1 + \frac{\kappa^{2}}{v \, n})}$ & $\bigg[\bar{y} + \frac{\kappa^{2}\log k}{n\, d} - \frac{d}{2}, \infty\bigg)$\\
    \bottomrule
  \end{tabular}%
  }
\end{table}
\endgroup Perhaps the simplest choice is a prior that does not depend on the
parameter value $\theta_0$ of the null hypothesis, such as a normal prior with
mean $m$ and variance $v$ (left column in Table~\ref{tab:normalinference}). The
hyperparameters $m$ and $v$ may be specified based on external data or based on
an alternative hypothesis of interest. For example, suppose that
$50$ students from the previous year had a mean test score of
$120$, which we now use to set the prior mean and variance to $m =
120$ and $v = 15^2/50$, see
Figure~\ref{fig:normal} for the resulting SC (orange). In this case, the MEE is
given by $\thatME = \bar{y} = 110$ with the support interval
centered around it. Due to the apparent conflict between the observed data and
the specified prior under the alternative, the $k = 1$ support interval spans a
wide range from $101.9$ to $118.1$,
indicating that average up to above average test scores are supported by the
data over the alternative.

The formulae in Table~\ref{tab:normalinference} (left column) show that as the
prior mean $m$ becomes closer to the observed data $y$, the evidence level
$\kME$ decreases and the support interval becomes narrower. This is because an
alternative closer to the data clearly has better predictive accuracy of the
data than an alternative further away, and thus fewer point null hypotheses can
outpredict it. Figure~\ref{fig:normal} illustrates this phenomenon with another
prior distributions (one with mean at the observed test score $\bar{y} =
110$, the blue SC). The SC has its mode still at the observed
mean test score but shows a far narrower $k = 1$ support interval from
$108.1$ to $111.9$ than the orange SC
with the mean $m = 120$ based on the previous year students.

Another approach to formulating a prior distribution for $\theta$ under the
alternative commonly suggested in `objective' Bayes theories is to center the
prior around the tested parameter value $\theta_{0}$ \citep{Jeffreys1961,
  Berger1987b, Kass1995b}. For example, one can specify a normal prior with mean
at the null value $\theta_{0}$ (middle column in
Table~\ref{tab:normalinference}). Thus, the resulting SC varies both the null
and the alternative, unlike the SC based on the `global' normal prior with fixed
mean $m$. As a result, the interpretation of the SC is different: For such a
`local' normal prior, the SC quantifies the support of parameter values over
alternative parameter values in a neighborhood around them. As for the global
normal prior, the MEE based on the local normal prior is given by $\thatME = y$
and support intervals are centered around it, but the associated, Bayes factor,
evidence level and support interval are different. Figure~\ref{fig:normal}
illustrates that when the mean $m$ of a global normal prior is too different
from the observed mean $\bar{y}$ (as in the case of the orange SC, where the
prior was specified based on the students from the previous year), the $k=1$
support interval based on the local prior with the same variance is narrower. On
the other hand, when the mean $m$ of the global prior is equal to the data (blue
SC), the support interval based on the local prior is wider.

The last prior in the right most column of Table~\ref{tab:normalinference}
represents a point prior shifted from the null value $\theta_{0}$ by $d > 0$.
The prior is again `local' in the sense that it is different for each tested
parameter value of the null hypothesis $\theta_{0}$, and as such encodes the
alternative hypothesis that the mean test score is greater than the tested
parameter value. However, this leads to an ever-increasing SC, see
Figure~\ref{fig:normal} for an illustration. As a result, the MEE and its
evidence level do not exist, while the support interval still exists but its
right limit extends to infinity. Although such a prior seems unrealistic, the
example demonstrates that a poorly chosen prior can lead to pathological
behavior of the resulting SC.

\subsection{Choice of the prior}
\label{sec:priorchoice}
While Bayes factors do not depend on the prior probabilities of their contrasted
hypothesis, they can be sensitive to the prior distributions specified for model
parameters under each hypothesis. Consequently, the parameter priors also have a
substantial impact on SC inferences, as the previous example showed. Moreover,
unlike posterior distributions, where the influence of the prior diminishes as
data accumulate, the sensitivity of Bayes factors to prior distributions
persists even with large samples \citep{Kass1995}.

This sensitivity is not inherently a weakness. It enables data analysts to
accurately quantify the support of parameter values over informative alternative
hypotheses when they are available \citep{Vanpaemel2010}, but poses a challenge
in their absence \citep{Liu2008, Williams2023}. Even when external information
exists, translating it into a prior distribution is often nontrivial. For
instance, knowing that a parameter lies within a certain range still leaves
considerable flexibility in the actual specification of the prior.

Several strategies have been proposed to address these challenges, each with
advantages and limitations. First, so-called `default' or `objective' priors aim
to yield Bayes factors with desirable properties, such as model selection
consistency, without incorporating prior knowledge \citep{Bayarri2012,
  Consonni2018}. These are often provided as defaults in software for computing
Bayes factors, such as the \texttt{BayesFactor} R package \citep{Morey2024} or
\texttt{JASP} \citep{Love2019}. Second, prior elicitation techniques can be used
to formally translate expert judgments into quantitative priors
\citep{OHagan2019, Stefan2022b, Mikkola2024}. Third, Bayes factor bounds
\citep{Berger1987, Sellke2001, Held2018} and sensitivity analysis
\citep{Kass1995, Sinharay2002, Franck2019} explore how Bayes factors change
across a class of priors to assess robustness of conclusions. Finally,
reverse-Bayes analyses identify the priors that lead to the Bayes factor
crossing a tipping point, allowing analysts to assess the plausibility of such a
prior in light of external knowledge \citep{Good1950,Held2021b}. As all of these
techniques are applicable to Bayes factors, they are also applicable to SCs.

As in other Bayes factor applications, SCs are only unambiguously defined if
priors for focus parameters are proper (i.e., integrate to one) under the
alternative $H_1$. Assigning improper or very vague priors can lead SCs to
diverge to infinity thereby proving overwhelming support for every tested
parameter value $\theta_0$, which is known as the `Jeffreys-Lindley paradox'
\citep{Robert2014, Wagenmakers2021a}. Priors for nuisance parameters may be
improper as long as the same prior is assigned under both the null $H_0$ and the
alternative $H_1$ so that arbitrary constants cancel out.

A general distinction can be made between \emph{global} and \emph{local} priors.
Global priors do not depend on the parameter value $\theta_{0}$ of the null
hypothesis. In contrast, local priors are defined relative to $\theta_0$. For
example, a local prior may be centered around $\theta_0$ or shifted away from
it, as the priors underlying the green and yellow SCs in
Figure~\ref{fig:normal}. This results in a different alternative hypothesis for
each tested parameter value, making the interpretation of the SC more intricate.
For a more natural interpretation, global priors may hence be preferred over
local priors. At the same time, local priors correspond to the typical use of
default Bayes factors, which is to center the prior around $\theta_{0}$, and
as such may be used in the same situations where default Bayes factors would be
used.

Given these considerations, SCs, just as Bayes factors, are especially useful if
informative priors based on prior data or external knowledege are available. If
such priors are lacking and default priors are used, it is advisable to report
sensitivity analyses for plausible ranges of priors or reverse-Bayes analysis,
to assess the robustness of the conclusions. A convenient visual sensitivity
analysis is, for example, to plot different SCs resulting from different prior
specifications, as shown in Figure~\ref{fig:normal}.

\subsection{Sequential analysis}
An attractive property of Bayesian inference is that it provides a coherent way
to analyze data that come in batches. That is, the same posterior distribution
is obtained regardless of whether all data are analyzed at once, or whether the
posterior distribution based on one batch is used as the prior for the other.

If we have two batches $y_{1}$ and $y_{2}$, the SC based on both batches is
\begin{align*}
  \BF_{01}(y_{1:2} ; \theta_{0})
  &= \BF_{01}(y_{1} ; \theta_{0}) \times \BF_{01}(y_{2} \mid y_{1} ; \theta_{0})
\end{align*}
where
\begin{align*}
  \BF_{01}(y_{2} \mid y_{1} ; \theta_{0})
  &= \frac{\int_{\Psi} p(y_{2} \mid \theta_{0}, \psi) \, p(\psi \mid y_{1}, H_{0}) \, \mathrm{d}\psi}{\int_{\Theta} \int_{\Psi} p(y_{2} \mid \theta, \psi) \, p(\theta, \psi \mid y_{1}, H_{1}) \, \mathrm{d}\psi  \, \mathrm{d}\theta}
\end{align*}
is the \emph{partial Bayes factor} obtained from using the posterior
distributions under the null $p(\psi \mid y_{1}, H_{0})$ and under the
alternative $p(\theta, \psi \mid y_{1}, H_{1})$ based on the first batch $y_{1}$
to compute the Bayes factor based on the second batch $y_{2}$
\citep[p.186]{OHagan2004}. This result generalizes to more than two batches by
\begin{align*}
  \BF_{01}(y_{1:n} ; \theta_{0})
  &= \BF_{01}(y_{1} ; \theta_{0}) \times \prod_{i=2}^{n} \BF_{01}(y_{i} \mid y_{1:(i-1)} ; \theta_{0}),
\end{align*}
that is, a SC based on all the available data can be obtained by multiplying
the SC based on the previous batches by the partial Bayes factor based on the
current batch. Thus, like ordinary Bayesian inference with posterior
distributions, SC inference is sequentially coherent.

\subsection{Asymptotic behaviour of the support curve}
It is of interest to understand the asymptotic behaviour of the SC, that is,
how does the SC (and quantities derived from it) behave as more data are
generated under a certain `true' hypothesis? It is well-known that Bayes factors
are consistent in the sense that when the data are generated under one of the
contrasted hypotheses, the Bayes factor tends to overwhelmingly favour that
hypothesis over the alternative as more data are generated, i.e., go to zero or
infinity, depending on the orientation of the Bayes factor \citep[see
e.g.,][]{Kass1992, Gelfand1994, Dawid2011}. Since the SC is nothing else than
the Bayes factor evaluated for various null hypotheses, this consistency
property carries over to the SC. That is, as more data are generated from the
true model with parameter $\theta_{*}$, the SC at $\theta_{0} = \theta_{*}$
will go to infinity, while the SC at $\theta_{0} \neq \theta_{*}$ will go to
zero.

As a concrete example where the distribution of the SC can be derived in
closed-form, consider again inference about a normal mean based on the data mean
$\bar{y}$ with distribution $\bar{Y} \mid \theta \sim \Nor(\theta,
\kappa^{2}/n)$, where $\kappa^{2}$ denotes a unit-variance and $n$ the sample
size. The logarithm of the SC based on a normal prior $\theta \mid H_{1} \sim
\Nor(m, v)$ can then be written as
\begin{align}
  \log \BF_{01}(\bar{Y};\theta_{0})
  &= \frac{1}{2}\left[\log\left(1 + \frac{n \, v}{\kappa^{2}}\right) +
 \frac{(\theta_{0} - m)^{2}}{v} -
    \left\{\bar{Y} - \frac{(\theta_{0} - m)\kappa^{2}}{n \, v} - \theta_{0}\right\}^{2}\frac{v \, n}{\kappa^{2}(v + \kappa^{2}/n)} \right].
  \label{eq:logbfnormal}
\end{align}
Hence, when data means are generated from $\bar{Y} \mid \theta_{*} \sim
\Nor(\theta_{*}, \kappa^{2}/n)$ with true mean $\theta_{*}$, we have that
\begin{align*}
    \left\{\bar{Y} - \frac{(\theta_{0} - m)\kappa^{2}}{n \, v} - \theta_{0}\right\}^{2} \frac{n}{\kappa^{2}} \sim \chi^{2}_{1,\lambda}
\end{align*}
with non-centrality parameter
$\lambda = n\left\{\theta_{*} - \frac{(\theta_{0} - m)\kappa^{2}}{n \, v} - \theta_{0}\right\}^{2}/\kappa^{2}.$
Thus, by rearranging terms in~\eqref{eq:logbfnormal}, we can compute the
probability that the Bayes factor is below some threshold $\gamma$ by
\begin{align}
  \label{eq:bfcdf}
  \Pr\{\BF_{01}(\bar{Y};\theta_{0}) \leq \gamma \mid \theta_{*}\} =
  1 - \Pr(\chi_{1,\lambda}^{2} \leq X)
\end{align}
with
\begin{align*}
  X = \left\{
  \log\left(1 + \frac{n \, v}{\kappa^{2}}\right) +
  \frac{(\theta_{0} - m)^{2}}{v} -
  2\log \gamma
  \right\} \left(1 + \frac{\kappa^{2}}{v\, n}\right).
\end{align*}

\begin{figure}[!tb]
\begin{knitrout}
\definecolor{shadecolor}{rgb}{0.969, 0.969, 0.969}\color{fgcolor}
\includegraphics[width=\maxwidth]{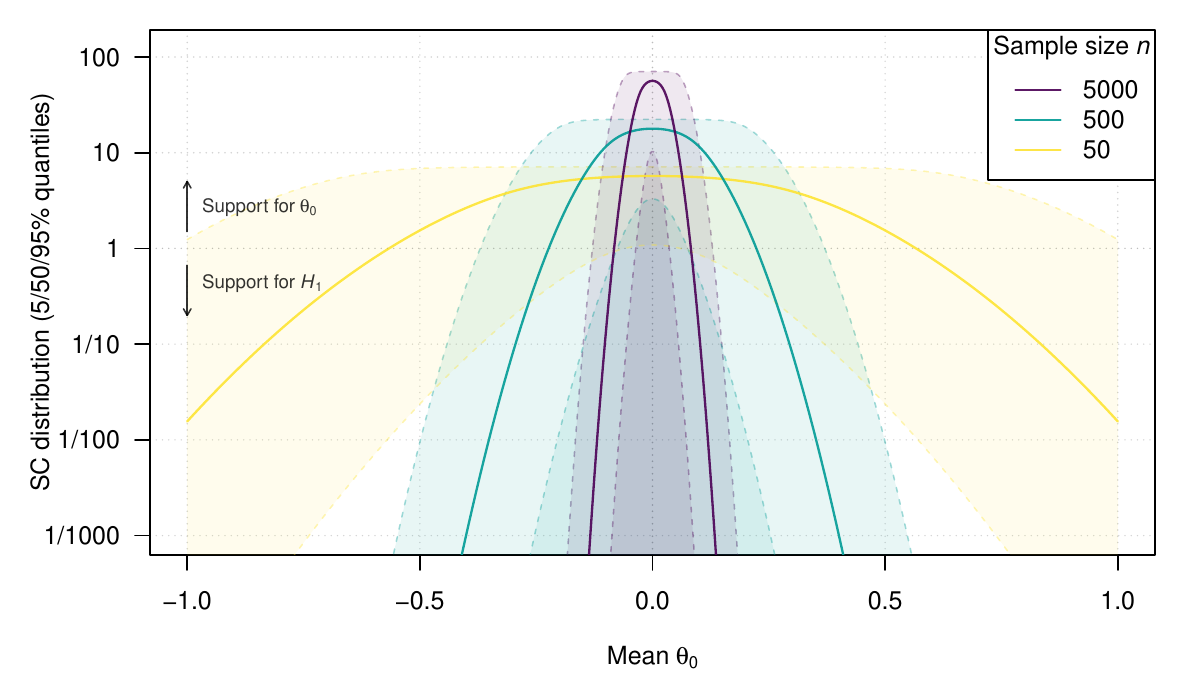} 
\end{knitrout}
\caption{The distribution of the SC, summarized by 5\% quantile (lower dashed
  line), 50\% quantile (solid line), 95\% quantile (upper dashed line), for
  different sample sizes $n$ (colors). A data model $\bar{Y} \mid \theta \sim
  \Nor(\theta, \kappa^{2}/n)$ is assumed with true mean $\theta_{*} =
  0$ and unit-variance $\kappa^{2} = 4$. The SC is
  based on a local normal prior $\theta \mid H_{1} \sim \Nor(\theta_{0}, v =
  4)$ assigned to $\theta$ under the alternative. Quantiles were
  computed by numerically inverting the cumulative distribution function of the
  Bayes factor~\eqref{eq:bfcdf}.}
\label{fig:SCdistribution}
\end{figure}

Figure~\ref{fig:SCdistribution} shows the distribution of the SC for different
sample sizes, a true mean of $\theta_{*} = 0$, a unit-variance of
$\kappa^{2} = 4$, and with a local normal prior with the same unit
variance \citep[a unit-information prior, see][]{Kass1995b} specified under the
alternative. We see that as the sample size increases, the distribution of the
SC at the true mean shifts toward larger values, indicating more evidence for
the true mean, as it should. On the other hand, the further away the SC is
evaluated from the true mean, the more its distribution shifts toward smaller
values, indicating increasing evidence for the alternative, as it should.

\section{Connection to other inference frameworks}
\label{sec:connections}

We will now explore connections of SC inference to other inference frameworks.

\subsection{Maximum integrated likelihood}

In typical situation where a division of the null's marginal likelihood
$p(y \mid H_0)$ by the alternative's marginal likelihood $p(y \mid H_1)$ does
not change the maximizer of the null's marginal likelihood $p(y \mid H_0)$, the
MEE can be obtained by maximizing $p(y \mid H_0)$ without reference to an
alternative $H_1$. This is, for instance, the case when a global prior (a prior
that does not depend on $\theta_{0}$) is assigned to $\theta$ under the
alternative, or also in the case of the local normal prior that is centered
around $\theta_{0}$ from the previous example. The MEE is then equivalent to the
maximizer of the \emph{integrated likelihood}
\begin{align*}
    \that_{\mathrm{MIL}} = \argmax_{\theta \in \Theta} \int_\Psi p(y \mid \theta, \psi) \, p(\psi \mid H_0) \, \mathrm{d}\psi,
\end{align*}
based on prior $p(\psi \mid H_0)$ assigned to the nuisance parameters
\citep[see e.g.,][]{Kalbfleisch1970, Basu1977, Berger1999, Royall1997,
  Severini2007}. When there are no nuisance parameters, the MEE reduces to the
ordinary maximum likelihood estimate.

To consider a concrete example, assume a sample of $n$ normal random variables
$Y_{1},\dots, Y_{n} \mid \theta, \sigma^{2} \overset{i.i.d.}{\sim} \Nor(\mu, \sigma^{2})$.
Suppose that $\sigma^{2}$ is the focus and $\mu$ the nuisance parameter, and
that an improper uniform prior $p(\mu \mid H_{0}) = 1$ is assigned to $\mu$. The
intergrated likelihood of an observed sample $y_{1}, \dots, y_{n}$ is then
\begin{align*}
  p(y_{1}, \dots, y_{n} \mid \sigma^{2})
  = (2\pi \sigma^{2})^{-(n - 1)/2} \, n^{-1/2} \, \exp\left\{\frac{-\sum_{i=1}^{n}(y_{i} - \bar{y})^{2}}{2\sigma^{2}}\right\}
\end{align*}
and maximizing it leads to the sample variance (REML) estimate of the variance
\begin{align*}
  \hat{\sigma}^{2}_{\mathrm{MIL}}
  = \frac{\sum_{i=1}^{n}(y_{i} - \bar{y})^{2}}{n - 1}.
\end{align*}
The same MEE is obtained when the prior $p(\mu, \sigma^{2} \mid H_{1})$ does not
depend on the value of the variance under $H_{0}$ as the denominator of the
Bayes factor is simply a multiplicative factor that does not change its maximum.
This shows that REML and MIL estimates can also be motivated from a Bayes factor
perspective which complements the well-established connections between REML
estimation and marginal posterior estimation based on flat priors for the
nuisance parameters \citep{Harville1974, Laird1982}. It is reassuring that
different methods produce the same estimate in these situations. However, the
important difference between these methods is the motivation and interpretation
of the resulting estimate -- the MEE represents a natural estimate for $\theta$
because it is the parameter value for which the data provide the most evidence
over an alternative hypothesis, while the (integrated) maximum likelihood
estimate is defined without reference to alternatives.

\subsection{Likelihoodist inference}
The likelihoodist school of statistical inference \citep{Barnard1949,
  Edwards1971, Royall1997, Blume2002} rejects the use of prior distributions to
formulate alternatives or to eliminate nuisance parameters, but it also shares
features with the SC paradigm. That is, if point priors are assigned to the
parameters, the Bayes factor reduces to a likelihood ratio which is the evidence
measure used by likelihoodists. For this reason, SC inferences correspond to
likelihoodist inferences if the Bayesian and likelihoodist agree on the used
point priors.

However, there is disagreement when it comes to the use of support sets. When
there are no nuisance parameters, likelihoodists define their support sets based
on the relative likelihood
\begin{align}
  L(\theta) = \frac{p(y \mid \theta)}{p(y \mid \that_{\textrm{ML}})}
  \label{eq:relLik}
\end{align}
where $\that_{\textrm{ML}}$ is the maximum likelihood estimate. For example,
\citet{Royall1997} recommended reporting the set of parameter values with
relative likelihood greater than $k = 1/8$ (at most `strong' evidence against
them) or $k = 1/32$ (at most `quite strong' evidence against them). From a
Bayesian perspective, using the observed maximum likelihood estimate as a prior
under the alternative seems to hardly represent genuine prior knowledge or an
alternative theory, but rather a cherry-picked alternative that gives to the
most biased assessment of support for the alternative \citep{Berger1987}.

\subsection{Frequentist inference}
The relative likelihood~\eqref{eq:relLik} serves as an important basis for
frequentist statistics since under the null hypothesis $-2\log L(\theta_{0})$
has an asymptotic chi-squared distribution with $\dim(\theta)$ degrees of
freedom. Frequentists thus also use relative likelihoods but merely as a test
statistic.

Another connection between frequentist and SC inference is given by the
`universal bound' \citep{Kerridge1963, Robbins1970, Royall1997, Sanborn2013},
which bounds the frequentist probability of obtaining misleading Bayesian
evidence. That is, when data are generated under $H_{0}: \theta = \theta_{0}$,
the probability of obtaining misleading evidence against the null at level $k$
(with a $\BF_{01} \leq k < 1$) is at most $k$ for any prior under the
alternative
\begin{align*}
  \Pr\{\BF_{01}(y) \leq k \mid H_{0}\} \leq k.
\end{align*}
This can be shown by noting that the expectation of $\BF_{10}$ under $H_0$ is 1
and applying Markov's inequality \citep[see e.g.,][]{Grunwald2024}. If there are
nuisance parameters, the bound holds only marginalized over the prior of the
nuisance parameters. For the bound to hold in a strict sense (i.e., for every
possible value of the nuisance parameter), special priors must be assigned to
them \citep{Hendriksen2021, Grunwald2024}.

The universal bound can thus be used to transform SCs into conservative
\textit{P}-values and confidence sets, e.g., a $k = 1/20$ support set obtained
from a SC corresponds to a 95\% conservative confidence set and $p =
\min\{\BF_{01}, 1\}$ corresponds to a conservative $P$-value. Remarkably, the
bound holds without adjustment even in `optional stopping' settings where data
collection is continuously monitored and stopped as soon as evidence against
$H_{0}$ is found \citep{Robbins1970}. However, it is important to note that
\textit{P}-values and confidence sets obtained in this way are usually much more
conservative than ordinary ones which are calibrated to have exact type I error
rate and coverage, respectively. Finally, if the data model is misspecified, the
bound is invalid.

\subsection{Bayesian inference}
\label{sec:bayesian}
The SC can, under certain conditions, be transformed into a Bayesian posterior
distribution. Specifically, assuming that the priors for the nuisance parameters
satisfy
\begin{align}
  \label{eq:SDcondition}
  p(\psi \mid H_{0}) = p(\psi \mid \theta = \theta_{0}, H_{1}),
\end{align}
the Bayes factor can be represented as the ratio of marginal posterior to prior
density under $H_1$ evaluated at the tested parameter value \citep[known as
  Savage-Dickey density ratio, see e.g.,][]{Dickey1971, Verdinelli1995,
  Wagenmakers2010}. Hence, the posterior under $H_1$ can be obtained by
multiplying the SC with the prior
\begin{align}
  \label{eq:posterior}
  p(\theta \mid y, H_{1})
  = \underbrace{\frac{p(y \mid \theta, H_{1})}{p(y \mid H_{1})}}_{= \BF_{01}(y; \theta)}
  \times p(\theta \mid H_{1}).
\end{align}
It is, however, important to emphasize that SCs based on priors under the
alternative that depend on the null (e.g., commonly used `local' normal or
Cauchy priors that are centered around $\theta_{0}$) cannot be transformed to a
genuine posterior distribution in this way, but multiplication with the prior
will result in a different posterior for every $\theta$.

Since, under certain regularity conditions, the posterior is asymptotically
normally distributed around the maximum likelihood estimate \citep[chapter
5.3]{Bernardo2000}, we can conclude that whenever these conditions are satisfied
and the SC has the Savage-Dickey density ratio
representation~\eqref{eq:posterior}, the SC is asymptotically given by the
asymptotic posterior normal density divided by the prior density, both evaluated
at $\theta_{0}$. The posterior, and hence also the SC, will become more
concentrated around the true parameter $\theta_{*}$ as more data are generated,
giving another intuition about the consistency property of Bayes factors.

The Savage-Dickey density ratio~\eqref{eq:posterior} also provides a convenient
way to compute SCs: One of the many programs for computing Bayesian posterior
distributions, such as \texttt{Stan} \citep{Carpenter2017} or \texttt{INLA}
\citep{Rue2009}, can be used to compute a posterior density, which can then be
divided by the prior density to obtain a SC. The caveat is again that this only
works for global priors under the alternative and provided that
condition~\eqref{eq:SDcondition} holds for the priors assigned to the nuisance
parameters.

The relationship between the posterior and the SC also exposes its connection
to another Bayesian inference quantity -- the \emph{relative belief ratio}
\begin{align}
  \label{eq:RB}
  \mathrm{RB}(\theta \mid H_{1}) = \underbrace{\frac{p(\theta \mid y, H_{1})}{p(\theta \mid H_{1})}}_{= \BF_{01}(y; \theta)},
\end{align}
see e.g., \citet{Evans2015}. This quantity is the updating factor of the prior
to the posterior density/mass function, and is related to the Bayes factor via
the aforementioned mentioned Savage-Dickey density ratio. An estimation and
testing framework centred on the relative belief ratio was developed by
\citet{Evans1997}. The parameter value that maximizes the relative belief ratio
was termed the \emph{least relative surprisal estimate}, later also referred to
as \emph{maximum relative belief estimate} \citep{Evans2015}. Clearly this
estimate is equivalent to the MEE whenever the SC and relative belief ratio
coincide. \citet{Evans1997} also defined a $\delta \times 100\%$ \emph{relative
surprise region}, which is the set of parameter values with $\delta \times
100\%$ posterior probability and with highest relative belief ratios among all
such sets. Similarly, \citet{Shalloway2014} defined an \emph{evidentiary
credible region} which is equivalent to the relative surprise region, but
motivated by information theory. While both are closely related to the support
set via the Savage-Dickey density ratio, they differ from the support set in
that they are defined by posterior probabilities and not by evidence, the
ordering induced by the relative belief ratio merely provides a rule to chose
among all credible sets \citep{Wagenmakers2020}. Thus, a relative surprise
region may contain parameter values that are not supported by the data. For this
reason, \citet{Evans2015} defined yet another type of region, a $k$
\emph{plausible region} which contains parameter values with a relative belief
ratio of at least $k$ and as such coincides with the $k$ support set whenever
the Savage-Dickey density representation applies to the SC and when a global
prior is chosen for $\theta$ under the alternative. Finally, Bayes factors and
relative belief ratios can also be seen as special cases of `Bayesian evidence
values' whenever the Savage-Dickey density ratio applies, so interval estimates
based on Bayesian evidence values also correspond to support intervals in these
cases \citep{Kelter2022}.


\section{Applications}
\label{sec:applications}

We will now illustrate SCs with the analysis of a binomial proportion
(Section~\ref{sec:binomial}), meta-analysis (Section~\ref{sec:metaanalysis}),
replication studies (Section~\ref{sec:replicationstudies}), and logistic
regression (Section~\ref{sec:logisticregression}).

\subsection{Binomial proportion}
\label{sec:binomial}

\citet{Bartos2023} conducted a study to test the hypothesis that fair coins tend
to land on the same side as they started slightly more often (with a probability
of about 0.51). This hypothesis was formulated by \citet{Diaconis2007} based on
a physical model of coin flipping. During the course of the study, 48
participants contributed to the collection of
$n = 350'757$ coin flips among
which \mbox{$y = 178'079$} landed
on the same side as they started.

We will now assume a binomial data model $Y \mid \theta \sim \Bin(n, \theta)$
and conduct inferences regarding the unknown probability $\theta$. In their
pre-registered analysis, \citet{Bartos2023} specified a truncated beta prior for
the probability $\theta$ under the alternative
($\theta \mid H_{1} \sim \Beta(a, b)_{[l, u]}$). Based on this prior, the Bayes
factor for testing $H_{0} \colon \theta = \theta_{0}$ against
$H_{1} \colon \theta \neq \theta_{0}$ is
\begin{align*}
  \BF_{01}(y ; \theta_{o})
  = \frac{\theta_{0}^{y} \, (1 - \theta_{0})^{n - y}}{\B(a + y, b + n - y)/\B(a, b)} \times
  \frac{I_{u}(a, b) - I_{l}(a, b)}{I_{u}(a + y, b + n - y) - I_{l}(a + y, b + n - y)}
\end{align*}
with the beta function
$\B(a, b) = \int_0^1 t^{a-1} \, (1 - t)^{b - 1} \, \mathrm{d}t$ and the
incomplete regularized beta function
$I_{x}(a, b) = \{\int_0^x t^{a-1} \, (1 - t)^{b - 1} \, \mathrm{d}t\} /\B(a, b)$.
Specifically, \citet{Bartos2023} assigned the hyperparameters
$a = 5100, b = 4900, l = 0.5, u = 1$ to instantiate an
alternative hypothesis that closely aligns with the theoretical prediction from
\citet{Diaconis2007} of a 0.51 probability with slight uncertainty around it.

\begin{figure}[!htb]
\begin{knitrout}
\definecolor{shadecolor}{rgb}{0.969, 0.969, 0.969}\color{fgcolor}
\includegraphics[width=\maxwidth]{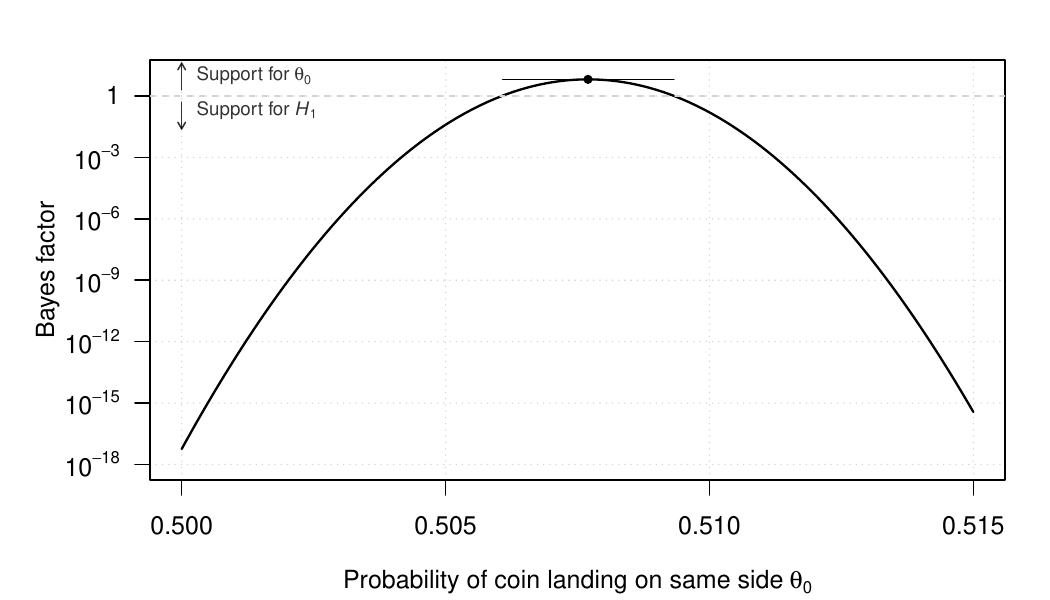} 
\end{knitrout}
\caption{Support curve analysis of data from \citet{Bartos2023}. Among
  $n = 350'757$ coin flips,
  \mbox{$y = 178'079$} landed on
  the same side as they started. A beta prior tightly concentrated around the
  theoretically predicted probability of 0.51 is assigned to the probability
  under the alternative
  (\mbox{$\theta \mid H_{1} \sim \Beta(5100, 4900)_{[0.5, 1]}$}).}
  \label{fig:bartosproportion}
\end{figure}

Figure~\ref{fig:bartosproportion} shows the resulting SC for a range of
probabilities from 0.5 to 0.515. Looking at the SC evaluated at $\theta = 0.5$,
we can see the finding reported by \citet{Bartos2023}: There is extreme evidence
($\BF_{01} = 1/(\ensuremath{1.76\times 10^{17}})$) against $\theta = 0.5$ and
in favour of the alternative concentrated around $\theta =0.51$. This result
hence provides decisive evidence for the theory from \citet{Diaconis2007} over
the hypothesis that coins tend to land on the same side with equal probability.
However, the SC framework permits further insights. For example, we can see
that all probability values up to about 0.504 and all values larger than 0.512
are decisively refuted by the data, each having an associated Bayes factor below
$10^{-3}$. Furthermore, the $k = 1$ support interval from 0.506 to 0.509 shows
the probability values that are better supported by the data than the specified
alternative, which excludes the theoretically predicted $\theta = 0.51$. The MEE
at $\hat{\theta}_{\mathrm{ME}} = 0.508$ is the best supported value, with
$\kME = 6.51$ indicating substantial evidence over the alternative concentrated
around 0.51.

Both the $k=1$ support interval and the MEE coincide with the 95\% central
credible interval and posterior mean based on a uniform prior distribution which
were reported by \citet{Bartos2023} alongside the Bayes factor for $\theta =
0.5$. The difference, however, is that the MEE, Bayes factor, and support
intervals are all coherently linked to the same SC based on the same prior and
data model, whereas the mix of Bayes factor, posterior mean, and credible
interval is not.

\subsection{Meta-analysis}
\label{sec:metaanalysis}
\begin{figure}[!phtb]

\begin{knitrout}
\definecolor{shadecolor}{rgb}{0.969, 0.969, 0.969}\color{fgcolor}
\includegraphics[width=\maxwidth]{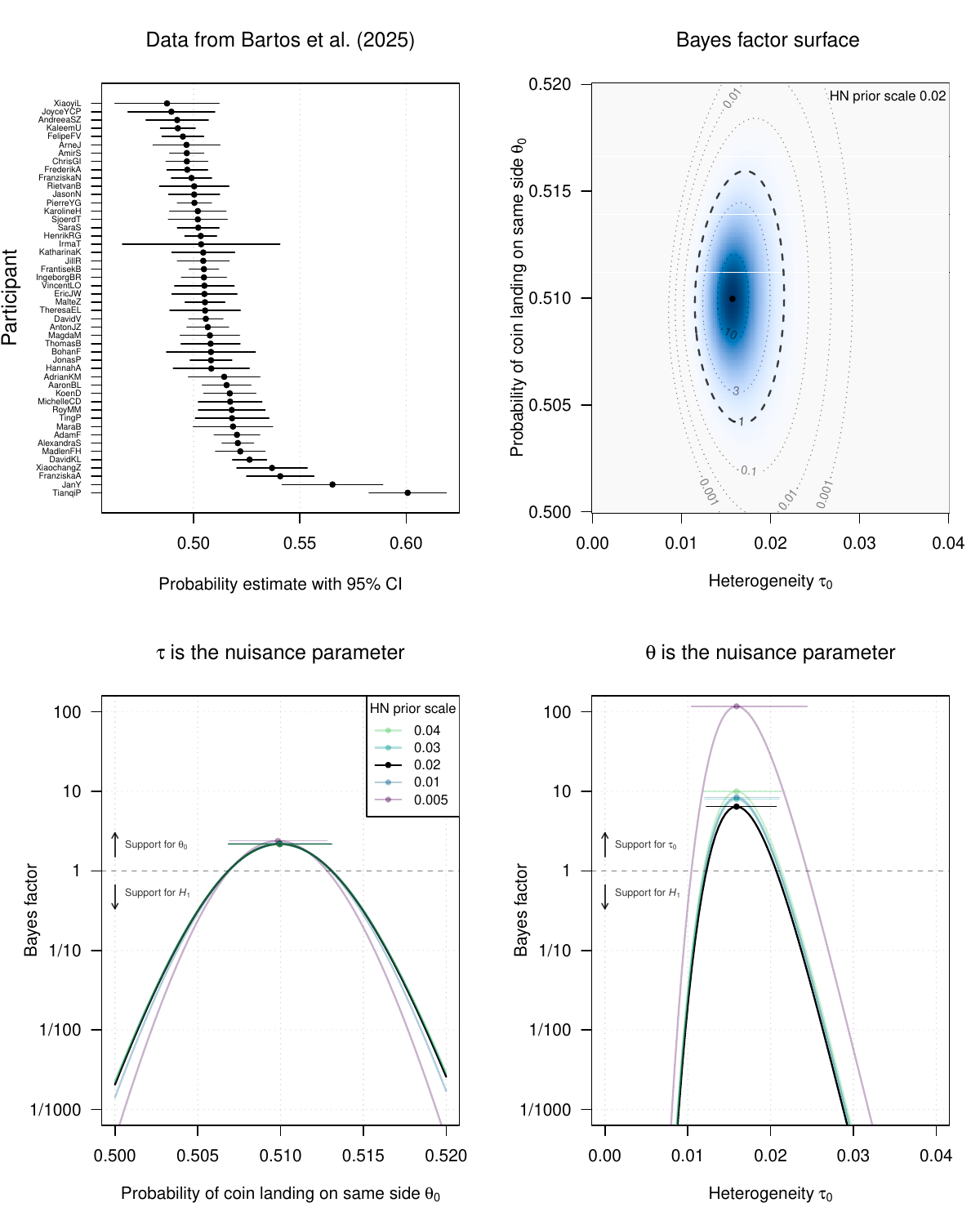} 
\end{knitrout}
\caption{Bayes factor analysis of coin flipping experiments from
  \citet{Bartos2023}, taking into account between-participant heterogeneity. The
  product of a truncated beta prior
  (\mbox{$\theta \mid H_{1} \sim \Beta(5100, 4900)_{[0.5, 1]}$})
  for $\theta$ and a half-normal prior with scale 0.02 for $\tau$ are
  assigned under the alternative $H_1$. The same priors are assumed when the
  parameters are nuisance parameters under $H_0$ (bottom plots). The bottom
  plots also show the SC for other scale parameters of the half-normal prior.}
\label{fig:bartosmeta}
\end{figure}

The previous analysis assumed that coin flips were independent among
participants and trials. The top left plot in Figure~\ref{fig:bartosmeta} shows
that this assumption seems violated as the estimated probabilities that a coin
lands on the same side for each of the $n = 48$ study participants are clearly
heterogeneous. This suggests that the analysis should be modified to account for
heterogeneity. In the following, we will therefore synthesize these estimates
while accounting for heterogeneity with a meta-analysis, as \citet{Bartos2023}
did.

Suppose we have $i = 1, \dots, n$ estimates $y_i$ with (assumed to be known)
standard errors $\sigma_i$. The estimates are assumed to be normally distributed
around a participant specific true probability $\theta_i$, i.e.,
\begin{align*}
    y_i \mid \theta_i, \sigma^2_i &\sim \Nor(\theta_i, \sigma^2_i) \\
    \theta_i \mid \theta, \tau &\sim \Nor(\theta, \tau^2).
\end{align*}
While in some applications, inferences about the participant specific
probabilities $\theta_i$ would also be of interest, these are consider nuisance
parameters in this analysis. Consequently, the distribution of an estimate is
marginalized over the participant probabilities distribution, leading to the
model
\begin{align*}
    y_i \mid \theta, \tau, \sigma^2_i &\sim \Nor(\theta, \sigma^2_i + \tau^2).
\end{align*}
There are two unknown parameters left, $\theta$ and $\tau$. The mean $\theta$
quantifies the average probability across participants, while the heterogeneity
standard deviation $\tau$ quantifies the heterogeneity of these probabilities.
The Bayes factor for testing $H_{0} \colon \theta = \theta_{0}, \tau = \tau_{0}$
against $H_{1} \colon \theta \neq \theta_{0}, \tau \neq \tau_{0}$ is then given
by
\begin{align*}
  \BF_{01}(y_{1}, \dots, y_{n}; \theta_{0}, \tau_{0})
  = \frac{\prod_{i}^{n} \Nor(y_{i} \mid \theta_{0}, \sigma^2_i + \tau_{0}^{2})}{
  \int_{0}^{\infty} \int_{0}^{1} \prod_{i}^{n} \Nor(y_{i} \mid \theta, \sigma^2_i + \tau^{2}) \,
  p(\theta,\tau \mid H_{1}) \, \mathrm{d}\theta \, \mathrm{d}\tau}
\end{align*}
with $\Nor(x \mid m, v)$ denoting the normal density with mean $m$ and variance
$v$ evaluated at $x$.


As in the previous analysis we assigned a
$\theta \mid H_{1} \sim \Beta(5100, 4900)_{[0.5, 1]}$
prior to the average probability $\theta$ under the alternative $H_1$. In
addition, we assigned a half-normal prior
$p(\tau \mid H_{1}) = \sqrt{2/\pi} \, \exp\{-\tau^{2}/(2 \, s^{2})\}/s$ to the
heterogeneity standard deviation $\tau$, and assumed it to be independent of
$\theta$. Half-normal priors are commonly used in meta-analysis due to their
simplicity and desirable properties such as nearly uniform behavior around zero
$\tau = 0$ \citep[see e.g.,][]{Rover2021}. We choose a scale $s = 0.02$
because the resulting prior gives
$95\%$ probability to $\tau$ values
smaller than 0.04, thus encoding the possibility of no heterogeneity (all true
participant probabilities are the same when $\tau = 0$) up to small amounts of
heterogeneity (the true participant probabilities differ by a few percentage
points). SCs for priors with smaller or larger scale parameters are also shown
in Figure~\ref{fig:bartosmeta} as sensitivity analyses.


The top-right plot in Figure~\ref{fig:bartosmeta} shows the SC in a
two-dimensional surface when both parameters are considered as focus parameters.
In contrast to the analysis that ignored between-participant heterogeneity, we
see that the MEE for the average probability
($\hat{\theta}_{\mathrm{ME}} = 0.51$) is now
consistent with the theoretical prediction of \citet{Diaconis2007}. In addition,
the MEE for the heterogeneity standard deviation
($\hat{\tau}_{\mathrm{ME}} = 0.016$) suggests
small but non-negligible heterogeneity. This MEE receives strong support over
the alternative ($\kME = 14$). The relatively
concentrated $k = 1$ support region indicates that probabilities from around
0.505 to 0.515 along with heterogeneity standard deviations from 0.012 to 0.021
are supported by the data over the alternative. Finally, the SC shows that
probabilities of $\theta = 0.5$ and no heterogeneity $\tau = 0$ are decisively
refuted by the data over the alternative
($\log \BF_{01} = \ensuremath{-1.81\times 10^{5}}$).


The two bottom plots in Figure~\ref{fig:bartosmeta} show SCs when either $\tau$
or $\theta$ is considered as nuisance parameter. In both cases, the same prior
as for the alternative $H_{1}$ was assigned to the corresponding nuisance
parameter under $H_0$. In addition, SCs for other choices of the scale parameter
of the half-normal prior were computed to assess the sensitivity of the results
to this choice. We see that the two marginal MEEs ($\hat{\theta}_{\mathrm{ME}} =
0.51$ and $\hat{\tau}_{\mathrm{ME}} =
0.016$) align with the joint MEEs, but their
evidence values ($\kME = 2.2$ and $\kME =
6.4$, respectively) indicate less support over the
alternative than for the joint one. Finally, looking at the colored SCs obtained
by changing the scale parameter of the half-normal prior assigned to $\tau$, we
see that the scale has little effect on inferences about the probability
$\theta$, but a more pronounced effect on inferences about $\tau$. For the
latter parameter, increasing the scale of the prior does not seem to change the
SC too much, while decreasing the scale to a value of $s = 0.005$ dramatically
increases the height of the SC, increasing the support of the MEE and
surrounding values over the alternative. This seems reasonable, since the data
show clear signs of heterogeneity, while a prior with such a small scale would
predict almost none.

In sum, the informative hypothesis tests carried out by \citet{Bartos2023} could
be embedded in a SC framework that additionally supplies them with compatible
point and interval estimates. This analysis suggests that, on average, coins
tend to land on the same side with probability $\theta = 0.51$ in accordance
with the hypothesis from \citet{Diaconis2007}. At the same time, there seems to
be non-negligible between-flipper heterogeneity, with a betwen-flipper standard
deviation of around one to two percent.

\subsection{Replication studies}
\label{sec:replicationstudies}

In a replication study, researchers repeat an original study as closely as
possible in order to assess whether consistent results can be obtained
\citep{NSF2019}. Various types of Bayes factor approaches have been proposed to
quantify the degree to which a replication study has replicated an original
study \citep{Verhagen2014, Ly2018, Harms2019, Pawel2022b, Pawel2023d}. A common
idea is that the posterior distribution of the unknown parameters based on the
data from the original study is used as the prior distribution in the analysis
of the replication data. If the replication data support this prior
distribution, this suggests replication success. We will now show how this idea
translates to analyzing replication studies with SCs.

Suppose that original and replication study provide effect estimates $y_{o}$ and
$y_{r}$ with standard errors $\sigma_{o}$ and $\sigma_{r}$, respectively. Each
is supposed to be normally distributed around a common underlying effect size
$\theta$ with (assumed to be known) variance equal to its squared standard
error, i.e., $y_{i} \mid \theta \sim \Nor(\theta, \sigma^{2}_{i})$ for
$i \in \{o, r\}$. A `replication SC' may then be obtained by using the
replication data to   contrast the null hypothesis
$H_{0} \colon \theta = \theta_{0}$ to the alternative
$H_{1} \colon \theta \sim \Nor(y_{o}, \sigma^{2}_{o})$, where the prior under
the alternative is the posterior distribution of $\theta$ based on the original
data and a flat prior for~$\theta$ \citep{Verhagen2014}. As such, the
replication Bayes factor represents a special case of the `partial Bayes factor`
\citep[p.186]{OHagan2004}. This leads to the following SC
\begin{align*}
  \BF_{01}(y_{r}; \theta_{0}) = \sqrt{1 + \frac{\sigma^{2}_{o}}{\sigma^{2}_{r}}}
  \, \exp\left[-\frac{1}{2}\left\{\frac{(y_{r} - \theta_{0})^{2}}{\sigma^{2}_{r}} - \frac{(y_{r} - y_{o})^{2}}{\sigma^{2}_{r} + \sigma^{2}_{o}}\right\}\right]
\end{align*}
with MEE at the replication effect estimate $\thatME = y_{r}$, evidence value
\begin{align*}
  \kME =  \sqrt{1 + \frac{\sigma^{2}_{o}}{\sigma^{2}_{r}}} \, \exp\left\{\frac{(y_{r} - y_{o})^{2}}{2(\sigma_{o}^{2} + \sigma^{2}_{r})}\right\},
\end{align*}
and $k$ support interval
\begin{align*}
  y_{r} \pm \sigma_{r} \sqrt{\log\left(1 + \frac{\sigma^{2}_{o}}{\sigma^2_{r}}\right) +
  \frac{(y_{r} - y_{o})^{2}}{\sigma^{2}_{r} + \sigma^{2}_{o}} - \log k^{2}}.
\end{align*}

We will now demonstrate application of the replication SC by reanalyzing data
from the replication project by \citet{Wagenmakers2016}. This project attempted
to replicate the original study from \citet{Strack1988} across
17 different study sites. The original study tested the
so-called `facial feedback hypothesis' and found that participants gave higher
funniness ratings to cartoons if they were smiling as opposed to showing
discontent (estimated mean difference of $0.82$ units on a
10-point Likert scale, with 95\% confidence interval from
$-0.05$ to $1.69$). The replications
were conducted across 17 different study sites, each producing a mean difference
effect estimate and confidence interval. In contrast to the original study, the
pooled replication mean difference was very close to zero (estimated mean
difference of $0.03$ with 95\% confidence interval from
$-0.11$ to $0.16$). We will now
assess the replicability of the original finding using replication Bayes factors
and the corresponding SCs.

\begin{figure}[!tb]
\begin{knitrout}
\definecolor{shadecolor}{rgb}{0.969, 0.969, 0.969}\color{fgcolor}
\includegraphics[width=\maxwidth]{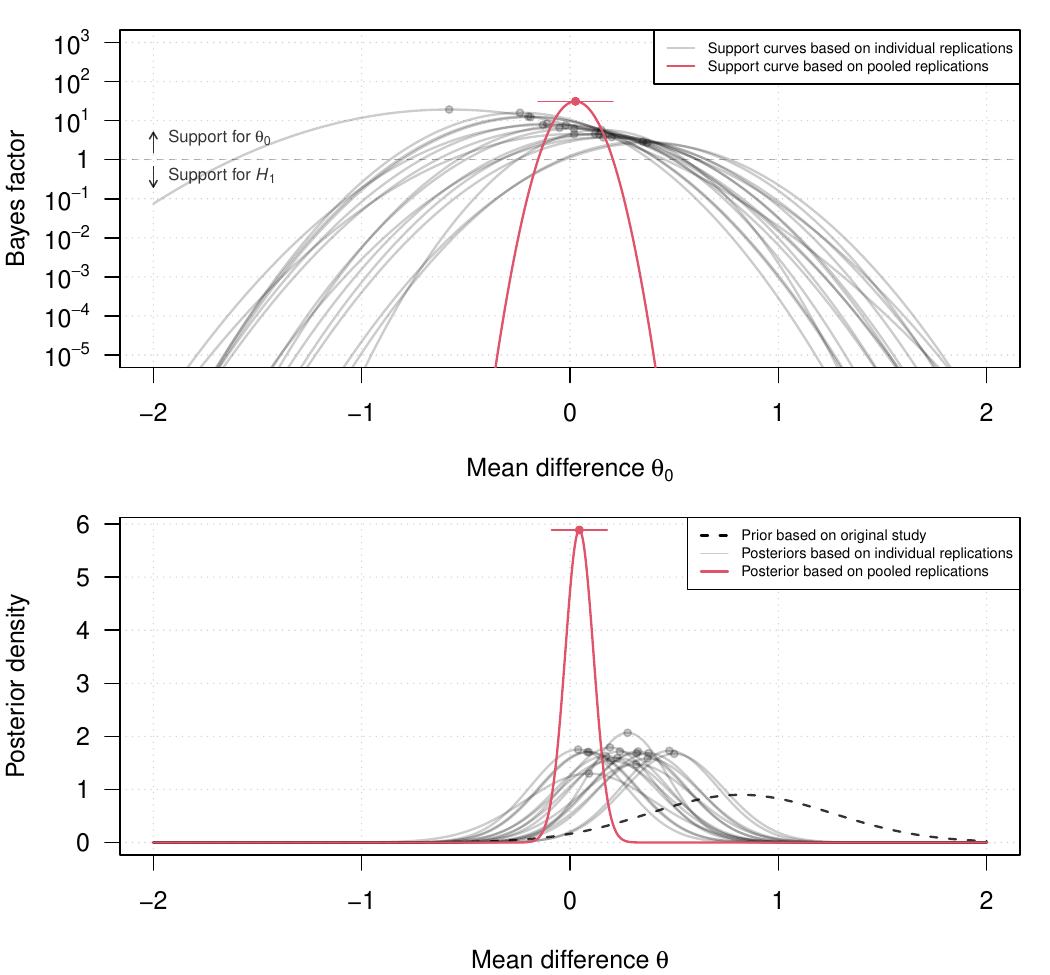} 
\end{knitrout}
\caption{Support curves with maximum evidence estimate and $k=1$ support
  interval (top) and posterior distribution with posterior mode and 95\% highest
  posterior density credible interval (bottom) for the pooled replication
  studies from the `facial feedback hypothesis' \citep{Strack1988,
    Wagenmakers2016}. The original study found an estimated standardized mean
  difference of $y_{o} = 0.82$ with standard error $\sigma_{o} =
  0.44$ which is used to formulate the prior distribution under
  the alternative $\theta \mid H_{1} \sim \Nor(y_{o}, \sigma^{2}_{o})$. The
  replication effect estimates are pooled via inverse variance weighting. The
  posterior is obtained by multiplying the SC with the prior density.}
\label{fig:replication}
\end{figure}

The top plot in Figure~\ref{fig:replication} shows the associated replication
SCs, MEE, and $k=1$ support interval. We see from the SC based on the
pooled replications that mean difference values close to zero receive more
evidence compared to the prior based on the original study
($\hat{\theta}_{\text{ME}} = 0.03$ with $k=1$ support
interval from $-0.15$ to $0.21$),
all of them being much smaller than the estimated mean difference from the
original study 0.82. Furthermore, the SC at the mean difference
of zero indicates strong evidence in favour of no difference over the prior based
on the original study ($\mathrm{BF}_{01} = 29$). Thus, the replication SC analysis suggests hardly any
replicability of the original facial feedback effect.

The bottom plot in Figure~\ref{fig:replication} illustrates the posterior
distributions, conveniently obtained by multiplying the SC by the prior
distribution based on the original data since this SC has a Savage-Dickey
density ratio representation. We can see that the $k=1$ support intervals from
the top plot are given by the set of effect sizes with posterior density larger
than the prior density. We can also see that for the pooled replications the
SC inferences mostly align with the posterior inferences, that is,
the maximum a posteriori estimate and 95\% credible interval are very close to
the MEE and $k=1$ support interval. This is because when pooling all the
replication data, the data are much more informative than the prior based on the
original study and thus overturn it. However, when comparing the individual
replications' SCs and posterior distributions we can see much larger
differences. The SCs peak at the corresponding replication effect estimates
while the posteriors peak at weighted averages of original and replication
estimates. For example, the replication with the smallest effect has a SC with
$\hat{\theta}_{\text{ME}} = -0.58$ and with $k= 1$
support interval from $-1.62$ to
$0.46$, whereas the corresponding posterior has its
mode at $\hat{\theta}_{\text{MAP}} = 0.09$ with
$95\%$ credible interval from $-0.51$ to
$0.69$. This apparent discrepancy reflects the fact
that SCs do not synthesize the data with the prior, as opposed to the posterior
analysis. Since in a replication study the interest is often to challenge the
original study, synthesis with the original data is arguably undesirable in the
analysis of replication studies. As such, a replication SC analysis, as
presented here, can be useful because it can clearly show whether or not there
are parameter values that are better supported than the prior based on the
original study.

\subsection{Logistic regression}
\label{sec:logisticregression}
To illustrate a more computationally demanding application of SCs, we consider
the criminological study by \citet{EjbyeErnst2023}. This study evaluated the
effect of a text-based light projection intervention (``\textit{It's illegal to
  buy drugs from street dealers}'' and a pictogram illustrating the same
message) on the presence of street drug dealers in Amsterdam's red-light
district. To this end, the researchers analyzed 765 one-minute segments of video
footage from two surveillance cameras installed in the red-light district with
respect to the presence/absence of street dealers before and after the
intervention, as well as other covariates (day of the week, camera, number of
people, and hours after 8 p.m.). Out of 765 segments, street dealers were
identified in 60. This small number of events makes it challenging to draw
conclusive inferences about the effect of the intervention, as we will see in
the following.

\begin{figure}[!htb]

\begin{knitrout}
\definecolor{shadecolor}{rgb}{0.969, 0.969, 0.969}\color{fgcolor}
\includegraphics[width=\maxwidth]{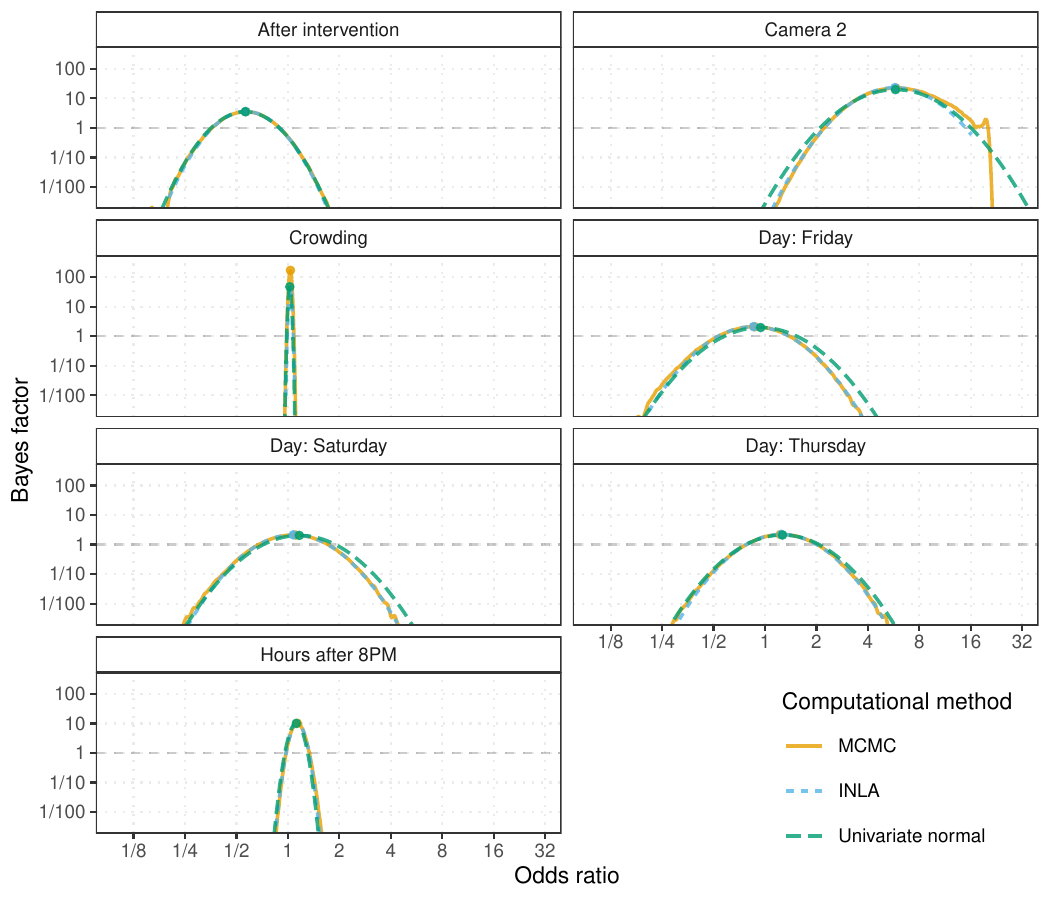} 
\end{knitrout}
\caption{Multiple logistic regression SC analysis of presence of street drug
  dealers in the Amsterdam red-light district before and after intervention
  \citep{EjbyeErnst2023}. Each plot shows the SC related to the exponentiated
  regression coefficient which can be interpreted as odds ratio. Independent,
  weakly informative $\Nor(0, 1/2)$ priors are assigned to the coefficients
  under the alternative $H_{1}$. All other coefficients were considered as
  nuisance parameters and the same priors assigned to them as under the
  alternative. Variables are binary indicators, except `Crowding' (the number of
  people at the beginning of a video observation) and `Hours after 8PM'.
  Wednesday is the reference category for the `Day' covariate.}
\label{fig:glm}
\end{figure}

Figure~\ref{fig:glm} shows the SCs related to a logistic regression analysis of
the data including an intercept term and a main effect for each covariate. Each
SC relates to the exponentiated regression coefficient, which can be interpreted
as the multiplicative change of odds of street dealer presence when increasing
the variable by one unit while keeping all other variables fixed. An improper
flat prior was assigned to the intercept under both the null and the
alternative, while independent $\Nor(0, 1/2)$ priors were assigned to the
coefficients under the alternative. This prior represents a weakly informative
alternative postulating that the median odds ratio is $1$ and that $95\%$ of
odds ratios are in between $1/4$ and $4$, representing a vague but realistic
range of odds ratios in observational data \citep{Greenland2006}. For the
analysis of each coefficient, all other coefficients were considered as nuisance
parameters with the same priors assigned to them under the null as under the
alternative. Note that even though we only look at the SC of a single parameter,
this approach does not ignore the nuisance parameters but averages the marginal
likelihood over their prior, thereby incorporating nuisance parameter
uncertainty.

SCs were computed in three ways: i) By first computing the marginal posterior
distribution for each coefficient from kernel smoothing of $1'000'000$ Markov chain Monte
Carlo (MCMC) samples (solid orange lines) computed with \texttt{Stan}
\citep{Carpenter2017} and then computing the SC via the Savage-Dickey density
ratio as explained in Section~\ref{sec:bayesian}. ii) By computing the marginal
posterior with integrated nested Laplace approximation (INLA; short-dashed blue
lines) via \texttt{INLA} \citep{Rue2009} and then computing the SC via the
Savage-Dickey density. iii) By estimating the parameters of the logistic model
first with maximum likelihood, and then using each estimated coefficient and its
standard error for a univariate normal analysis similar to
Section~\ref{sec:replicationstudies}, ignoring the nuisance parameters in the SC
analysis (long-dashed green lines). As a result, the MEEs from the univariate
normal analysis correspond to maximum likelihood estimates, while the MEEs from
the MCMC and INLA analyses correspond to integrated maximum likelihood
estimates.

The MCMC analysis took the longest of the three (around 5 minutes to run),
followed by the INLA analysis (about a second to run), followed by the
univariate analysis (almost instantaneous). We can see that the SCs based on
MCMC and INLA (with default settings) may have inaccuracies or cannot be
computed in the outer regions of the SC, since these are regions where the
posterior density is close to zero. We can also see that the univariate normal
SC agrees reasonably well with the MCMC and INLA SCs in most cases, with the
exception of the `Camera 2' coefficient. In this case, the MCMC and INLA SCs end
abruptly around OR = 16, because no larger MCMC samples were observed or because
the INLA algorithm returned a posterior density of zero. In this case, the
normal approximation SC could be used cautiously as an extrapolation of the MCMC
and INLA SCs.

Concering the effect of the intervention, the SC (based on MCMC) for the
variable `After intervention (top left panel) has its mode at
$\widehat{\text{OR}}_{\mathrm{ME}} = 1/1.8$ indicating a negative
association between the text-based intervention and the presence of street
dealers, yet this parameter value receives only moderate support over the
alternative ($\kME = 3.5$) and the corresponding $k = 1$ support interval
spans the range from a noticeable association
($\text{OR}=1/2.8$) up to hardly any association ($\text{OR}
= 1/1.1$). This suggests some negative association between
the intervention and the presence of street dealers, although the extent of the
association being relatively unclear.

Among the remaining variables, the SC related to `Camera 2' clearly suggests a
higher prevalence of street dealers at the location of camera 2 compared to
camera 1 ($\widehat{\text{OR}}_{\mathrm{ME}} = 5.8$ with $\kME =
22.2$ and
$k = 1$ support interval from $\text{OR}=2.2$ to $\text{OR} =
17.1$) while the `Crowding' SC suggests a small but
positive association between the number of people in a video sequence and the
presence of street dealers ($\widehat{\text{OR}}_{\mathrm{ME}} =
1.04$ with $\kME = 169.3$ and $k = 1$ support interval from
$\text{OR}=1$ to $\text{OR} = 1.1$). On the other hand,
the SCs related to the day of the week and hours after 8 p.m. are largely
undiagnostic about whether or not the variables exhibit negative or possitive
associations with the presence of street dealers, possibly due to the sparse
nature of the data.

In sum, this example demonstrated how SCs can be applied to more complicated
models such as logistic regression, and how they can be obtained from
general-purpose programs for computing posterior distributions such as
\texttt{Stan} or \texttt{INLA}.

\section{Discussion}
\label{sec:discussion}
We showed how Bayes factors can be used for parameter estimation, extending
their traditional use cases of hypothesis testing and model comparison. We also
linked these ideas to the overarching concept of support curves (SCs),
which are Bayes factor analogues of \textit{P}-value functions, and are likewise
particularly useful for reporting of analysis results. This provides data
analysts with a unified framework for statistical inference that is distinct
from conventional frequentist and Bayesian approaches: While a \textit{P}-value
function can only quantify evidence \emph{against} parameter values
\citep{Greenland2023}, SCs allow us to quantify evidence \emph{in favour} of
parameter values over the alternative. Moreover, if a SC diagnoses absence of
evidence, data analysts can continue to collect data without worrying about
multiplicity issues. Like ordinary Bayesian inference, SC inference uses the
Bayesian evidence calculus, but without synthesizing data and prior. When point
priors are assigned, Bayes factors become likelihood ratios, so SC inference
aligns with likelihoodist inference, but when there are nuisance parameters,
SCs include a natural way to eliminate them via marginalization over a prior.

Like the likelihoodist and Neyman-Pearson paradigms of statistical inference, SC
inference requires the formulation of alternative hypotheses. For this reason,
SCs are particularly valuable in contexts where prior data or theories are
available to formulate alternative hypotheses. For example, SCs (under the name
of `\textit{K} ratio`) have been used by the large-scale NANOGrav collaboration
to quantify the evidence for new physics theories against the established
Standard Model \citep{Afzal2023}. Similarly, our application of SCs to the coin
flipping experiment from \citet{Bartos2023} or the replication studies from
\citet{Wagenmakers2016} was relatively straightforward as the prior distribution
for the parameters under the alternative could be specified through the a priori
specified hypotheses and the data from the original study, respectively. In
cases where there are no clear alternative hypotheses, data analysts may use SCs
based on `weakly informative' \citep{Gelman2009} or `default' prior
distributions \citep[e.g., unit-information priors, see][]{Kass1995b}, but
should acknowledge this limitation and report sensitivity analyses (e.g., SCs
for different prior distributions). Another possibility is to base SC inference
on Bayes factor bounds \citep{Berger1987, Sellke2001, Held2018}, which give a
bound on the maximum evidence against parameter values, but at the cost of
losing the ability to quantify evidence \emph{in favour} of parameter values
\citep{Pawel2023}.

We have focused on situations where null models are nested within the
alternative model. However, SC inference could also be applied to more general
model comparison where the null is not nested in the alternative. In this case,
the interpretation of the SC is more intricate as the comparison does not only
involve an additional parameter but an entirely different model. Moreover,
computing the SC via the Savage-Dickey density ratio is not possible anymore.

Where under their control, data analysts should design experiments and studies
so that conclusive inferences can be drawn from the data collected, including
SC inferences. Future research needs to investigate how experiments need to be
designed to enable conclusive inference with SCs. For example, one may design
an experiment so that the expected width of a support interval is sufficiently
narrow, or so that the expected evidence level for the MEE is sufficiently
large. Finally, calculating SCs can be challenging, as our logistic regression
example showed. For example, if a SC is computed via the Savage-Dickey density
ratio from a posterior distribution computed by MCMC, the SC may be imprecise
at the tails of the posterior, even with millions of samples. Future work may
focus on developing more efficient techniques for computing SCs in such
settings.

Bayesian, likelihoodist, or predictive reasoning may all motivate the Bayes
factor as a natural tool for quantifying the relative evidence or support of
competing hypotheses. Nevertheless, neither the Bayes factor nor any other
measure of statistical evidence is infallible or suitable for all purposes. In
fact, there is empirical evidence that researchers sometimes misuse or
misinterpret Bayes factors. For example, Bayes factors are sometimes erronously
interpreted as absolute rather than relative evidence, or as posterior odds
rather than updating factors \citep{Wong2022, Tendeiro2024}. Bayes factors by
construction do not take into account the prior probabilities of their
contrasted hypotheses, so they may indicate strong support for a hypothesis even
though this hypothesis would still remain unlikely when combined with its prior
probability \citep{Lavine1999, Good2001}. Any type of statistical inference can
lead to distorted scientific inferences if used in a bright-line fashion without
consideration of contextual factors \citep{Goodman2016, Greenland2023}. We
believe that SCs are useful in this regard because they shift the focus from
finding evidence against a single null hypothesis to making gradual and
quantitative inferences.

\section*{Acknowledgments}
We thank \citet{EjbyeErnst2023} and \citet{Bartos2023} for openly sharing their
data. We thank \anonymize{Riko Kelter, Andrew Fowlie, Eric-Jan Wagenmakers,
  František Bartoš, Leonhard Held, Małgorzata Roos, and Sander Greenland} for
valuable comments on drafts of the manuscript. We thank Yang Liu, Jonathan
Williams, and another anonymous reviewer for helpful and constructive comments
that substantially improved paper. The acknowledgment of these individuals does
not imply their endorsement of the paper.

\section*{Conflict of interest}
We declare no conflict of interest.

\section*{Software and data}
The data from \citet{EjbyeErnst2023} where obtained from the
`\texttt{data.RData}' file available at \url{https://osf.io/nb56d}. The data
from \citet{Wagenmakers2016} were manually extracted from Figure 4 in their
paper. The standard errors were recomputed by dividing the difference of the
upper and the lower confidence interval bounds by twice the 97.5\% quantile of
the standard normal distribution. The data from \citet{Bartos2023} were obtained
from the \texttt{dat.bartos2023} data set included in the \texttt{metadat} R
package \citep{White2023}. The code and data to reproduce our analyses is openly
available at \anonymize{\url{https://github.com/SamCH93/BFF}}. A snapshot of the
repository at the time of writing is available at
\anonymize{\url{https://doi.org/10.5281/zenodo.10817311}}. We used the
statistical programming language R version 4.4.1 (2024-06-14) for
analyses \citep{R} along with the \texttt{brms} \citep{Burkner2021} and
\texttt{INLA} \citep{Rue2009} packages for the computation of posterior
distributions.

{\small
\begin{spacing}{1.25}
\setlength{\bibsep}{4pt plus 0.3ex}
\bibliographystyle{apalikedoiurl}
\bibliography{bibliography}
\end{spacing}
}

\end{document}